\documentstyle[preprint,aps]{revtex}
\begin{document}
\draft

\title{Fractional dimensional Hilbert spaces and Haldane's exclusion statistics
}
\author{K.N. Ilinski$^{1,2}$\cite{Ilinski} and J.M.F. Gunn$^{1}$\cite{Gunn}}
\address{$^{1}$ School of Physics and Space Research, University of Birmingham,
Birmingham B15 2TT, United Kingdom.}
\address{$^{2}$ Institute of Spectroscopy, Russian Academy of Sciences,
Troitsk, Moscow region, 142092, Russian Federation.}

\maketitle

\begin{abstract}
We examine the notion of
Haldane's dimension and the corresponding statistics in a probabilistic
spirit. Motivated by the example of dimensional-regularization we
define the dimension of a space as the trace of a diagonal `unit operator',
where the diagonal matrix elements are not, in general, unity but are
probabilities to place the system into a given state. These
probabilities are uniquely defined by the rules of Haldane's
statistics.
We calculate the second virial coefficient for our system and demonstrate
agreement with Murthy and Shankar's calculation. The partition function for an
ideal gas of the particles, a state-counting procedure,
the entropy and a distribution function
for the particles are investigated using our probabilistic definition.
We compare our results with previous calculations of exclusion statistics.
\hfill hep-th/9503233\break
\end{abstract}

There has been considerable recent interest in a paper by Haldane \cite{Hald}
generalising the Pauli exclusion principle and introducing a combinatorial
expression for the entropy of particles which were generalisations of bosons
and fermions. Haldane initially defines a quantity $d(N)$,
which we will term the {\it Haldane dimension}, which is the dimension of
the one-particle Hilbert space associated with the $N$-th particle,
keeping the coordinates of the other $N-1$ particles fixed.
The statistical parameter, $g$, of a particle (or `$g$-on') is defined by
(where we add $m$ particles)
\begin{equation}
g = - \frac{d(N+m) - d(N)}{m}
\label{g}
\end{equation}
and the conditions of homogeneity on $N$ and $m$ are imposed. In addition
the system is confined to a finite region where the number $K$ of
independent single-particle states is finite and fixed.
Here the usual Bose and Fermi ideal gases have
$g=0$ for Bose case (i.e. $d(N)$ does not depend on $N$)
and $g=1$ for Fermi case -- that is the dimension is reduced by one for each
added fermion, which is the usual Pauli principle.

Haldane also introduced a combinatorial expression (which we denote the
Haldane-Wu \cite{Hald,Wu} counting procedure) for the number of ways,
$W$, to place $N$ $g$-ons into $K$ single--particle states. Then
\begin{equation}
W = \frac{(d(N)+N-1)!}{(d(N)-1)! N!}  \qquad d(N)=K-g(N-1) ,
\label{W}
\end{equation}
which was subsequently used by many authors \cite{MS1,WB,W2,Ha,Is,Raj} to
describe thermodynamical properties of $g$-ons. In particular Bernard and
Wu \cite{WB} and Murthy and Shankar \cite{MS2} showed that the behaviour of the
excitations in
the Calogero--Sutherland model is consistent with Eqn. (\ref{W})
\cite{Wu} for $g$-ons, with fractional $g$, in general.

In this Letter we argue that these two ideas introduced
by Haldane are {\it distinct}, in that the most natural combinatorial
expression for $W$, using the quantity $d(N)$, is {\it not} (\ref{W}).
This leads us to a definition of the dimension of a Hilbert space which
is {\it fractional} and is related in a natural manner to particles having
a fractional $g$. We construct the statistical mechanics for such
particles and show that many features agree with Wu's statistical
mechanics, however the agreement is not complete.

The argument for the expression for $W$ in (\ref{W}) is illuminated by
initially remembering the fermionic case, writing (\ref{W}) in the form
\begin{equation}
W_{F} = K (K-1) \cdots (K-N+1) / N!
\nonumber
\end{equation}
where the $i$th bracket in the numerator is the number of possibilities to
insert the $i$th particle in the system when all particles are distinguishable
and then correcting the answer with $N!$ in the denominator. By analogy,
expression (\ref{W}) can
be rewritten replacing the factor of the {\it initial}
number of available states, $K$, in fermionic expression by an
effective number of allowed states, $K+(1-g)(N-1)$. Thus the space of available
single-particle states {\it swells} before
the particles are added.
This contradicts the assumption of fixing the size of the system which leads
to a fixed number of single-particle states. From the definition of the Haldane
dimension, $d(N)$, we
would have na\"{\i}vely expected a result of the form
\begin{equation}
W_{0} = K (K -g)\cdots(K+(1-g)(N-1)-N+1) / N!
\label{S}
\end{equation}
where the number of available
states decreases proportionally to $g$ as the particles are added. This
expression has the drawback that it does not yield the Bose limit as $g \to 0$,
although is correct for integer $g\ge 1$.

It is very striking to compare the prediction of (\ref{W}) for the case $g=2$
and straightforward calculation for $K$ single-particle states and
$N$-particle sector. For example with $K=10$ and $N=3$
Haldane-Wu procedure gives $W=8!/(3!5!) =56$ and Eqn. (\ref{S}) gives $W_0=(10
\cdot 8 \cdot 6)/3!=80$ and the difference
is observable. It is easy to see that, for general large $K$ and $N$, the
following expansion for the deviation of $\ln W$ from
straightforward counting exists:
\begin{equation}
\ln W_{0} - \ln W = \frac{1}{2}\frac{N^2}{K} + O\left(\frac{N^3}{K^2}\right) \
{}.
\nonumber
\end{equation}
This deviation is
important for cases where the occupation numbers are not small.
The above discussion may be summarised by stating that Haldane's
definition of the fractional dimension and the Haldane-Wu state-counting
procedure are not consistent.

We will now derive a $W$ which is more in keeping with the notion of the
Haldane dimension. We will initially derive the result using one single
particle state and then generalise to the case with many states. The basis is a
{\it probabilistic} interpretation of a fractional Haldane dimension, and its
implementation as a fractional--dimension Hilbert space. Probability enters
since an attempt to add a particles to a particular single-particle state
succeeds only with a certain probability.

Consider the single-particle state to be initially empty.
We assume that the statistics only start to enter at the two-particle
level, so the probability of adding the first particle is unity:
$p_1=1$. However the next particle may only be added with probability
$p_2$. Then if we consider a large number, $M$, of copies of our system and
perform the
trials with each of them we find approximately $p_2 M$ doubly-occupied
states so that on average only a fraction $p_2$ of the particle is on each
site. We interpret this as the fractional dimension of the subspace associated
with
double occupation on the site and {\it define} the Haldane dimension to be
$d(2) = p_2$.
In general we have that $p_N$ is the probability of adding the $N$th
particle, given that there are already $N-1$ particles there
(i.e. the probability is {\it conditional}), and $p_N=d(N)$.
{}From this we may deduce the relation of $p_n$ to $g$ using (\ref{g}):
\begin{equation}
d(N)\equiv p_n = 1-(n-1)g \quad n\ge 1
\label{Single}
\end{equation}
In this context the
original Pauli principle `There are no double-occupied states'
can be reformulated as `The probability to construct a double-occupied state is
equal to zero'. We should note that if $g=1/m$, then there are no difficulties
with negative probabilities (as $p_{m+1}$ vanishes), and we shall restrict
ourselves to those cases~\cite{note}.

The geometric implementation of this is by analogy with
dimensional-regularization where the dimension of a
space is defined as the trace of a diagonal `unit operator' (or weight) where
the diagonal matrix
elements are not unity in general but are probabilities to find a system in
the state corresponding to the matrix element. With this definition we may
relate the probabilities used to relate the Haldane definition to the diagonal
weights, $q_N$, since the probability of sequentially adding $N$ particles to
the vacuum is
\begin{equation}
q_N = \prod_{n=1}^{N} p_n = \prod_{n=1}^{N}(1-[n-1]g) \ .
\label{q}
\end{equation}
Then the dimension ({\it not} the Haldane dimension,
as it is not a conditional probability) associated with the
$N$-particle subspace is $q_N$, and the total dimension of the many-particle
Hilbert space is the sum $\sum_{n=0}^\infty q_N$,
with the convention that $q_0 =1$. The weight, ${\mathcal I} $, is:
\begin{equation}
{\mathcal I}  = \mbox{diag}(q_{0},q_{1},q_2\cdots )
\end{equation}
and in the many single-particle state case
\begin{equation}
{\mathcal I} ^{(K)} = \sum_{n_{1},\cdots,n_{K}=0}^{\infty}
\alpha_{n_{1},\cdots,n_{K}}
|n_{1},\cdots,n_{K}\rangle\langle n_{K},\cdots,n_{1}| \ .
\label{Id}
\end{equation}
where $K$ is the number of states.
We may now define the partition function, and hence thermal
expectation values, of an arbitrary physical variable $\hat{O}$:
\begin{equation}
Z=\mbox{Tr\thinspace }  ({\mathcal I} ^{(K)} \thinspace e^{-\beta H}) \qquad
\mbox{or} \qquad
\langle\hat{O}\rangle=\mbox{Tr\thinspace }  ({\mathcal I} ^{(K)} \thinspace
e^{-\beta H} \hat{O})
\label{Z}
\end{equation}
where the Hamiltonian $H$ is
$
H = \sum_{i=1}^K \epsilon _{i} n_{i}
$
by analogy with a bose or fermi ideal gas and is independent of statistics. It
is
the weight, ${\mathcal I} ^{(K)}$, which completely defines the exclusion
statistics of the particles. In Eqn. (\ref{Id}) $|n_{1},\cdots,n_{K}\rangle$ is
the state with $n_{1}$ particles in
the first {\it single-particle} state, $n_{2}$ at the second and so on and
$\alpha_{n_{1},\cdots,n_{K}}$ is the probability to find this state.
It is obvious that the expression (\ref{Z}) gives right answers for
fermions ($\alpha_{n_{1},\cdots,n_{K}}=1$ for $n_{1}\&\cdots\&n_{K} \leq 1$ and
zero otherwise) and bosons ($\alpha_{n_{1},\cdots,n_{K}}=1$ for any
$n_{1},\cdots,n_{K}$) and there is no contradiction in using this expression
for statistics of an intermediate nature. Moreover we can say that
Hilbert space of the theory is constructed if we define the operator ${\mathcal
I} ^{(K)}$
and use the equalities (\ref{Z}). The mathematical
details will be discussed elsewhere.

We can now interpret $\alpha_{n_{1},\cdots,n_{K}}$ as the dimension of
the subspace spanned by the vector $|n_{1},\cdots,n_{K}\rangle$. Indeed,
usually dimension of a subspace $S$ can be defined as a trace of unit operator
on
the subspace
\begin{equation}
\mbox{dim}\ S = \mbox{Tr\thinspace } ({\mathcal I} ^{(K)}|_{S})
\end{equation}
and by analogy with the definition in
dimensional regularization, where $d-\epsilon = \sum_{i} \delta _{i}^{i}$.
The full dimension of the $N$-particle subspace of the space of states is then
given by the formula:
\begin{equation}
W_{0}= \mbox{Tr\thinspace }  ({\mathcal I} ^{(K)}|_{n_{1}+\cdots+n_{K}=N})
\label{W01}
\end{equation}
which we will use presently for the state-counting procedure. Then Haldane's
dimension $d(N)$ of $N$-th particle subspace
with arbitrary fixed $N-1$-particle substate
$|n_{1},\cdots,n_{K}\rangle,n_{1}+\cdots+n_{K}=N-1$ is described then by
relation:
\begin{equation}
d(N)= \sum_{l=1}^{k}
\frac{\alpha_{n_{1},\cdots,n_{l}+1,\cdots,n_{K}}}
{\alpha_{n_{1},\cdots,n_{K}}} \delta_{n_{1}+\cdots+n_{K},N-1} \ .
\label{dalph}
\end{equation}
Let us turn to the concrete choice of the
probabilities $\alpha_{n_{1},\cdots,n_{K}}$ and relate the result for one
single-particle state, (\ref{Single}), to the many state case. Although the
presence of many indices on the $\alpha$'s might indicate considerable choice,
in fact there is a
{\it single} self-consistent way to define $\alpha_{n_{1},\cdots,n_{K}}$ such
that the definition of the $N$-th particle dimension $d(N)$
(\ref{dalph}) actually gives the Haldane's conjecture $K-g(N-1)$ for
$d(N)$ and the Hilbert space of the system with $K$ degrees of freedom is
factorized as a product of Hilbert spaces corresponding to each degree of
freedom (this property together with the Hamiltonian
of an ideal gas of the particles
will lead to the factorisation of the partition function
and other physical quantities).

To show this we will use of the assumption about the tensor product nature
of the full Hilbert space,  which immediately gives the operator ${\mathcal I}
^{(K)}$ for
the full system as a tensor product of the operators $\{{\mathcal I}
_{i}\}_{i=1}^{K}$
for each state:
\begin{equation}
{\mathcal I} ^{(K)} = {\mathcal I} _{1}\otimes {\mathcal I} _{2} \otimes \cdots
\otimes {\mathcal I} _{K} \ .
\label{TP}
\end{equation}
The last relation is equivalent, using Eqn. (\ref{q}) to the following
expression for
probabilities $\alpha_{n_{1},\cdots,n_{K}}$:
\begin{equation}
\alpha_{n_{1},\cdots,n_{K}}=
\prod_{i=1}^{K} \left\{ (1-g)(1-2g)\cdots(1-(n_{i}-1)g)\right\} \ .
\label{alph}
\end{equation}
This leads, using (\ref{dalph}), to the required equality for $d(N)$:
\begin{equation}
d(N)= \sum_{l=1}^{k}
(1-g n_{l}) = K - g N \ .
\label{Hal}
\end{equation}
One of the main advances of the paper~\cite{MS1} was the generalisation
of the Haldane's definition, which originally was formulated for a
finite-dimensional systems, to the case of an infinite-dimensional Hilbert
space. Then the
following expression for the statistical parameter $g$ was obtained:
\begin{equation}
\frac{1}{2} - g = \lim_{K\rightarrow \infty} \frac{K}{N(N-1)}
\left[N! \frac{W}{K^{N}} - 1\right] \ .
\label{MS}
\end{equation}
and $W$ is given by eq.(\ref{W}). Now, replacing the original $W$ by the
`regulated definition of the dimension of the Hilbert space'
$\tilde{W}=\lim_{\beta \rightarrow 0} Z_{N}
=\lim_{\beta \rightarrow 0} \mbox{Tr\thinspace } (e^{-\beta H_{N}})$
and substituting it into the eq.(\ref{MS}) the following generalisation of
the Haldane's definition was proposed:
\begin{equation}
\frac{1}{2} - g =
\lim_{\beta \rightarrow 0}
\frac{Z_{1}}{N(N-1)}
\left[N! \frac{Z_{N}}{Z_{1}^{N}} - 1\right] \ .
\label{MS1}
\end{equation}
This definition (\ref{MS1})
was used in the paper~\cite{MS1} to connect $g$-ons with anyons and
excitations in the Luttinger model~\cite{note1}. However there is a very strong
assumption here that it is possible to calculate the original fractional
dimension, $W$, as a trace of the statistical operator. Thus the
results of the paper \cite{MS1} correspond rather to the systems which can
be described via traces (\ref{W01}) rather than the original (those defined in
Eqn. (\ref{g})) $g$-ons. Our paper is an
attempt to fill this breach and describe the original $g$-ons in terms of
traces:
we have already shown that our fractional Hilbert space
definition agrees with Haldane's expression for $N$-th
particle dimension $d(N)$ (\ref{Hal}). On the other hand we have shown that our
definitions lead to the correct results if we use the formula (\ref{MS1}) (so
probabilistic $g$-ons are $g$-ons
in the definition of Murthy and Shankar \cite{MS1}).

More precisely it is possible to check
that $W_{0}$ contains exactly the same
contributions proportional to $(1-g)^n$ as $W$ up to order
$O(N^{n+1}/K^{n+1})$, providing the same description of double-occupied
states (in particular this is reflected in the identity of the second virial
coefficients $B_{2} = \frac{1}{2}(g-\frac{1}{2})$).
So we can conclude that
the  behaviour of the systems at sufficiently low density (high enough,
however, that statistical corrections begin play the role) governed
by the state-counting dimension (\ref{W}) and (\ref{W01}) are equivalent.
Unfortunately we cannot to say the same about high order
corrections.

The factorisation of the statistical operator (\ref{TP}) implies a
factorisation of
the partition function of the system:
$
Z = \prod _{i=1}^{K} Z_{i}  \ ,
$
where the function $Z_{i}$ is the partition function of the single state
with index $i$:
$
Z_{i} = \mbox{Tr\thinspace } ({\mathcal I} ^{(K)}_{i} \cdot e^{-\beta
(\epsilon_{i} - \mu)n_{i}}) \ .
$
For the case of $g=1/m$ last expression can be expanded as
\begin{equation}
Z_{i}=\sum_{n=0}^{m} \ \ {}^{m}C_{n} \ \ n! \ \ \frac{e^{-n \beta
(\epsilon_{i}-\mu)}}{m^{n}} \ .
\label{Zi}
\end{equation}
It is obvious that the formula (and hence all statistical quantities)
interpolates
between the Fermi ($g=m=1$) and Bose ($g=0,m=\infty$) cases.

The distribution
function $n(\beta ,\mu) \equiv \langle N/K\rangle$
(where all single-particle states have energy $\epsilon$)
may be calculated to be:
\begin{equation}
n(\beta ,\mu) = \frac{
\sum_{n=1}^{m} \ {}^{m}C_{n} \ \ (n! n/m^{n}) \
e^{-n \beta (\epsilon_{i}-\mu)}}{
\sum_{n=0}^{m} \ {}^{m}C_{n} \ (n!/{m^{n}}) \ e^{-n \beta
(\epsilon_{i}-\mu)}} \ .
\label{n}
\end{equation}
Statistical
effects manifest themselves at low temperatures;
in the low
temperature limit, using (\ref{n}), the function $n(\beta ,\mu)$ for our
$g$-ons
behaves analogously to Wu's distribution function \cite{Wu}.  At low enough,
but finite, temperatures and energy, $\epsilon$, above
the Fermi level (i.e. $\xi = e^{\beta (\epsilon-\mu)}$ is very big and $m\ge
2$) we find the following
expansion for the distribution function $n(\beta ,\mu) = (1/\xi)(1 + [m-2]/[\xi
m] + \cdots )$. Above the Fermi level the behaviour of the distribution
function
(\ref{n}) and Wu's is similar (this similarity has the same
origin as the coincidence of second virial coefficients).

The situation crucially changes if we consider the behaviour of the
distribution functions below the Fermi level at low but finite temperatures
(i.e. $\xi $ is a small but finite parameter). From Wu's calculations we
immediately obtain the following approximate form for the distribution:
$n(\beta ,\mu)  = 1/m - \xi ^{m} + \cdots$. In contrast with this the
asymptotics for distribution function (\ref{n})
contains terms linear in $\xi$ : $n(\beta ,\mu) = 1/m - \xi/m + \cdots$. So our
distribution
function has distinct behaviour compared to Wu's below the Fermi
level; however the behaviour of our result may be more realistic.

Now let us return to the state-counting procedure for probabilistic $g$-ons.
For convenience we once more put all energies $\epsilon_{i}=\epsilon$ and
we rewrite the formulae for the dimension of $N$-particle subspace of the full
Hilbert space (\ref{W01}) in the form
without constrained summation
which then can be treated as a Fourier integral:
\begin{equation}
W_{0}= \frac{1}{2\pi}
\int _{0}^{2\pi } d\phi \sum_{n_{1},\cdots ,n_{K}=0}^{\infty}
\alpha_{n_{1},\cdots ,n_{K}} \cdot e^{i (n_{1}+\cdots +n_{K}-N)\phi} \ .
\nonumber
\end{equation}
Last equality and the factorisation property (\ref{alph}) for the
coefficients $\alpha_{n_{1},\cdots ,n_{K}}$ allow us to perform the summation
on $n_{1},\cdots ,n_{K}$ and obtain the expression for $W_{0}$ with only a
single integral on the auxiliary variable $\phi$:
\begin{equation}
W_{0}= \frac{1}{2\pi} \int _{0}^{2\pi } d\phi \ \
z(e^{i \phi})^K \cdot e^{-i N\phi} \ .
\label{int}
\end{equation}
where the function $z(e^{i \phi})$ coincides with the single state partition
function on imaginary energy and chemical potential ($\beta (\mu- \epsilon)
\rightarrow i \phi $).
The eq.(\ref{int}) substitutes in the theory of probabilistic $g$-ons
the state-counting Haldane-Wu expression (\ref{W}).

Wu and Bernard \cite{WB} state that systems which can be solved
by the Thermodynamical Bethe ansatz (TBA) can be considered as systems of
ideal gas of the
particles obeying Haldane's exclusion statistics.
Let us stress again that these systems {\it do not obey}
Haldane's exclusion statistics formulated in the terms of the $N$-th particle
dimension (in contradiction to previous work) but
indeed {\it are} described by
Haldane-Wu state-counting procedure with the formula (\ref{W}). Following
\cite{WB} we have a fermionic rule for the
calculation, but the number of places available for
occupation in the `system' (`cell') depends linearly on the number of particles
in the
system. Nevertheless, the procedure differs from Haldane's definition. Indeed,
unused (`free') places cannot be occupied by additional particles because they
change in the next ($N+1$) particle sector. This
can be observed, for example, in the Calogero model: in the $N$-particle sector
we can construct a motif using the notions of pseudo-particles and holes
(for the connection with exclusion statistics see \cite{Ha}). These holes
represent accessible orbitals but they cannot be filled by the $N+1$-th
particle as is required by Haldane's definition. This is not the only point
of difference: the procedures are different, as discussed at the begining of
this Letter.

In conclusion,
in this Letter we treated the notion of
Haldane's dimension and the corresponding statistics in a probabilistic
spirit, defining the dimension of a space as the trace of a diagonal `unit
operator'
where the matrix elements are the probabilities to place the system in a given
state. (The
probabilities are uniquely defined by the rules of Haldane's
statistics.) The Hilbert space
for particles with exclusion statistics was defined using this `unit operator',
allowing the construction of a second quantised formulation of the
theory \cite{13}. Contrary to statements in earlier works,
systems solvable by TBA are {\it not} an ideal gas of exclusion statistical
particles but are described by a different deformation of statistics,
based on the Haldane-Wu state-counting procedure.

We wish to thank A.S.Stepanenko and I.V.Lerner for useful discussions.
We thank the UK EPSRC for support under Grant number GR/J35221. This work was
also partially supported (K.N.I) by the International
Science Foundation, Grant N R4T000, the Russian Fund of Fundamental
Investigations, Grant N 94-02-03712, INTAS-939 and Euler stipend of German
Mathematical Society.

{\bf Addendum}: After this work was finished we saw reference \cite{Poly},
where the author examined the microscopic origin of the Haldane-Wu state
counting procedure. similarly to this work the notion of statistics was
considered in a probabilistic manner. He assumed that a single level may be
occupied by any number of particles, and each occupancy is associated with an a
priori  probability. These probabilities are determined by enforcing
consistency with the state counting procedure. However the distinctions with
this work are: no use was made of the definition of the Haldane dimension,
there was no construction of a Hilbert space and the a proiri probabilities may
be negative.  As we noted above the Haldane dimension and the Haldane-Wu state
counting procedures are distinct; the foundations of this work and reference
\cite{Poly} are distinct in the same manner.


\begin{thebibliography}{99}
\bibitem{note}
Although in this paper we deal with the simple expression for statistical
parameter $g=1/m$ the consideration of other value of $g$ is also
important. They should appear, for example, if we consider $t-J$ type
Hamiltonians to model interaction effects with finite Coulomb interaction
with $g<2$ and in Calogero type models with $g=p/q$.
\bibitem{note1}
There are also exact results on anyons in strong magnetic fields.
Thermodynamical quantities were obtained \cite{Ouvry} and display the
features of Haldane-Wu statistics.
\bibitem{13}
 This is especially important for the aim to model different
solid state systems on the basis of exclusion statistics such as
generalisation of the $t-J$ model (see also Ref.\cite{BS,W2}) and others.
\end{thebibliography}
\end{document}